\documentclass[3p, twocolumn]{elsarticle}

\usepackage{amsmath}
\usepackage{amssymb}
\usepackage{graphicx}
\usepackage{dcolumn}
\usepackage{bm,bbm}
\usepackage{color}
\usepackage{placeins}
\usepackage{multirow}
\usepackage{hyperref}
\usepackage{setspace}

\begin{document}
\onehalfspacing


\title{Quantitative schlieren imaging of a laser-ionized plasma channel in atomic vapor using symbolic regression}%

\author[1]{G. Demeter\corref{cor}}

\address[1]{HUN-REN Wigner Research Centre for Physics, Budapest, Hungary}

\cortext[cor]{email: \texttt{demeter.gabor@wigner.hun-ren.hu}}


\date{\today}

\begin{abstract}
The plasma channel of the AWAKE plasma wakefield acceleration experiment is created by laser ionization of rubidium vapor in a 10~m long vapor source.
Verifying the properties of the channel --- the radius of its fully ionized core and the width of its boundary sheath --- that are vital for ensuring good-quality wakefields is important as is validating numerical models of ionizing pulse propagation.
Near-resonant schlieren imaging can detect the plasma channel in the low-density vapor, but there is no direct inversion to obtain channel parameters from the images.
We use symbolic regression to find compact analytical formulas that connect measurable properties of the schlieren signal to plasma profile parameters.
Using databases of simulated signals generated with three different plasma edge-profile families, we obtain 
formulas with one and two fitting constants to predict the equivalent channel radius with an accuracy below twice the camera pixel size and formulas to predict the sheath width.
Re-evaluating previous measurements with these formulas, we confirm a theoretically predicted power-law relationship between the channel radius and the energy of the ionizing laser pulse.
The formulas can be adapted to new experimental conditions by refitting the constants on a modest set of new samples, providing a lightweight, interpretable alternative to evaluation with deep neural networks.
\end{abstract}

\maketitle

\section{Introduction}
\label{intro}

Schlieren imaging is a sensitive optical technique for visualizing spatially varying refractive index fields in transparent media \cite{Schlierenbook}.
A collimated probe beam is deflected where transverse refractive index gradients are present; a mask placed at the focal plane of a lens blocks the undeflected beam and passes the deflected portion. The second lens thus images only this deflected portion onto a camera.
Originally developed for aerodynamics and fluid dynamics, where it has been used to visualize shock waves, turbulent flows, and heat transfer \cite{Schlierenbook}, the technique has also found application in the study of laser-produced plasmas \cite{Iwase1998,Clayton1998,Honda2000,Veloso2006,Batani2019}, and has been tested for imaging atomic excitation \cite{Bachmann2018} and laser-generated plasma columns \cite{Demeter2024} in rubidium vapor.

The AWAKE experiment at CERN is a proton-driven plasma wakefield accelerator, aiming to accelerate electron bunches to hundreds GeV energy in a single plasma stage \cite{awake,Caldwell2016,Gschwendtner2016,Adli2018,Gschwendtner2022}.
The plasma channel required for wakefield generation is created by the photoionization of rubidium vapor along a $\sim$10 m long vapor source by a focused, short-pulse laser.
Because the homogeneity of the plasma density in a long channel of sufficiently large radius is essential for the quality of the proton-bunch self-modulation that drives the wakefield, as well as for the particle acceleration process, it is important to verify the properties of the plasma column. In particular, the radius of the channel core within which the plasma density is constant, and the width of the plasma sheath (the transition region at the plasma boundary) are important parameters. Numerical simulations of the ionizing pulse propagation \cite{Demeter2019,Demeter2021,Demeter2025} make predictions about these quantities that should be compared with experimental measurements to verify the numerical model. In turn, the verified numerical code can be used to predict the length limits of plasma channel creation in a given ionizing laser setup. 

When an intense laser pulse ionizes a gas or vapor, it changes the refractive index of the medium along its path, so the plasma channel it creates can be imaged by a transverse probe beam that samples this refractive index change across the channel cross section.
Such measurements are complicated by the low density of the rubidium vapor (of order $10^{14}$--$10^{15}$ cm$^{-3}$), which makes the plasma optically thin and its direct imaging challenging. Near-resonant schlieren imaging, that uses a probe beam tuned close to the rubidium D$_2$ resonance line, exploits the anomalous dispersion of the residual neutral vapor to obtain a large refractive index signal even at these low densities, making the plasma column detectable as a local reduction in neutral-atom density \cite{Bachmann2018,Demeter2024}.

The key challenge is then to extract \emph{quantitative} plasma channel parameters from the observed schlieren image. Because the schlieren signal depends on the phase integral of the probe through the entire column cross-section, modified by the spatial filtering of the mask, a direct analytical inversion is not possible.
Previous approaches to this inverse problem in the context of AWAKE have included training deep neural networks (DNNs) on large databases  of simulated schlieren signals \cite{Biro2023}.
While DNNs can predict plasma parameters with high accuracy, they require construction of very large training databases  ($\sim 10^6$ samples), expensive training by high-power GPU clusters, and full retraining whenever experimental parameters such as probe wavelength or vapor density change considerably.
A second approach \cite{Demeter2024} assumed a single Gaussian-edge plasma profile and identified, from numerically computed sample sets, that signal peak properties are well approximated by bilinear functions of the plasma core radius and sheath width; this yields a compact description with eight fitting constants per set of experimental parameters.

Symbolic regression is a machine learning technique that searches the space of analytical expressions --- most often with evolutionary algorithms --- for compact formulas that describe a dataset, balancing prediction accuracy against formula complexity \cite{Schmidt2009,Udrescu2020,Cranmer2023}.
In contrast to DNNs, which encode the learned input--output relationship in an opaque set of millions of network weights, symbolic regression yields explicit closed-form expressions that can be inspected, interpreted physically, and used in error propagation.
Because the class of candidate models is much more constrained, it typically requires substantially smaller training datasets than DNNs, at the price of being practical only for a modest number of input variables.

In this paper we take a more general and systematic approach to the schlieren inverse problem than we employed before, using symbolic regression.
We characterize the plasma column by two profile-agnostic parameters: the equivalent radius $R_{eq}$, defined through the area under the radial ionization probability profile, and a parameter characterizing the width of the plasma sheath.
We generate databases using a mixture of three distinct edge profile shapes (Gaussian, sigmoid, and cosine) and a random set of parameters governing the plasma core location, radius and sheath layer width.
Symbolic regression then discovers compact analytical formulas, with a small number of free constants, that link observed schlieren signal properties directly to these plasma parameters.
Once determined for a given set of experimental conditions, the formulas for the plasma radius can be adapted to new conditions (e.g. a different vapor density or probe wavelength) by refitting those constants alone on a modest database of new samples.
We describe the sample generation procedure, the signal analysis strategy, the resulting formulas, and discuss their accuracy, limitations, and advantages relative to previous approaches.

\section{Schlieren imaging of the plasma column}
\label{sec_setup}

The schlieren setup used for AWAKE plasma channel imaging is described in detail in \cite{Demeter2024}, and we only summarize the aspects most relevant to the present analysis.
A probe beam at $\lambda_p = 780.311$~nm (in vacuum), tuned close to the rubidium D$_2$ resonance at $\lambda_{D2} = 780.241$~nm, traverses the vapor source transversely.
Near the resonance, the anomalous dispersion of the neutral rubidium vapor contributes a refractive index change $\delta n \sim 10^{-4}$--$10^{-3}$, which is three to four orders of magnitude larger than the plasma dispersion contribution at the same density.
When the ionizing laser creates plasma, the local population of the rubidium ground state is reduced, lowering the refractive index in the plasma region and imprinting a spatially dependent phase shift on the probe beam.
The choice of probe wavelength is a compromise: the probe must be tuned close enough to resonance to obtain sufficient refractive index sensitivity, but far enough to avoid excessive resonant absorption in the vapor.
The optimal detuning also depends on the vapor density, in part through the collisional broadening of the atomic line \cite{vanLange2020}.

\begin{figure}[htb]
\centerline{\includegraphics[width=0.45\textwidth]{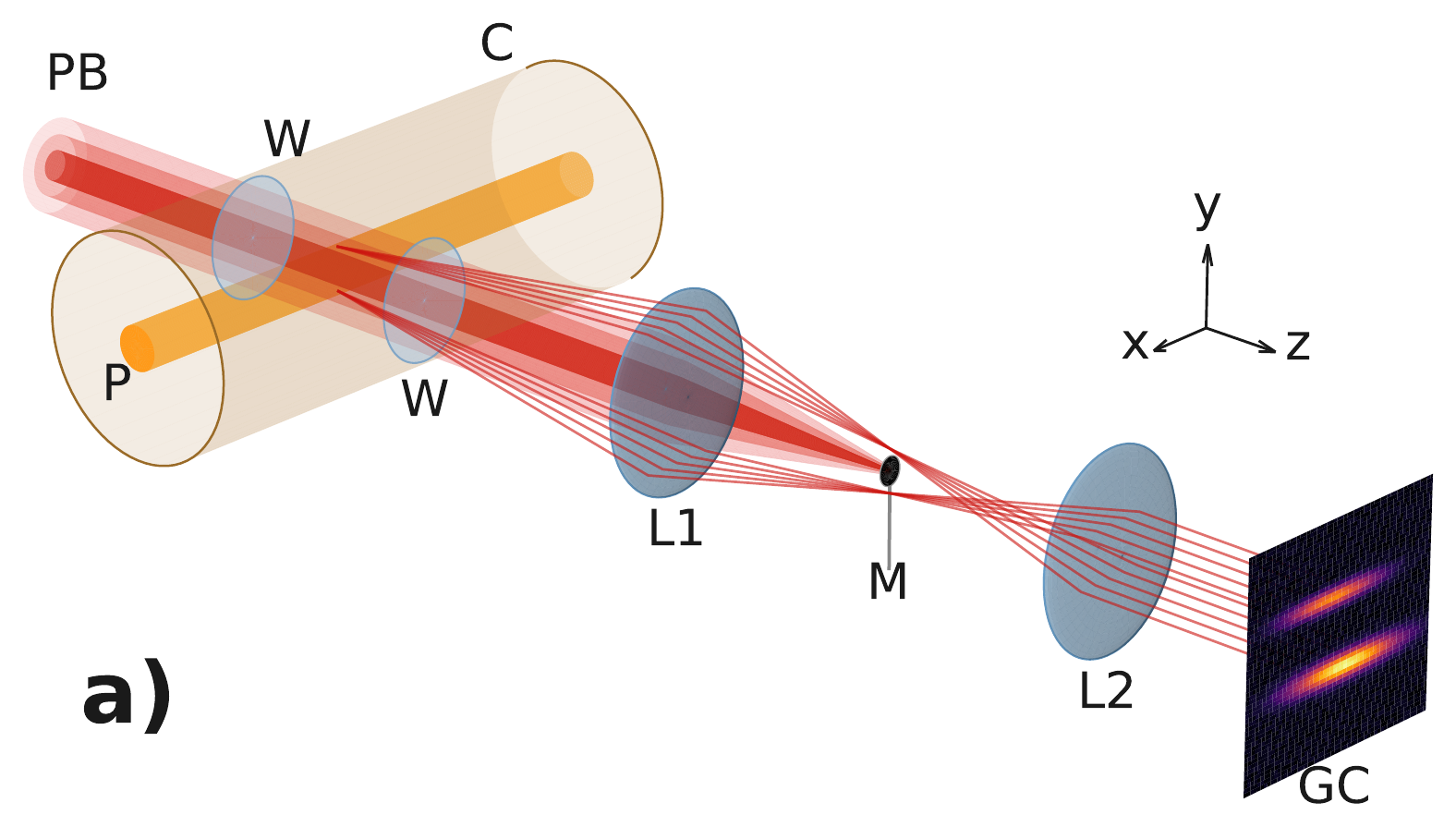}}
\centerline{\includegraphics[width=0.35\textwidth]{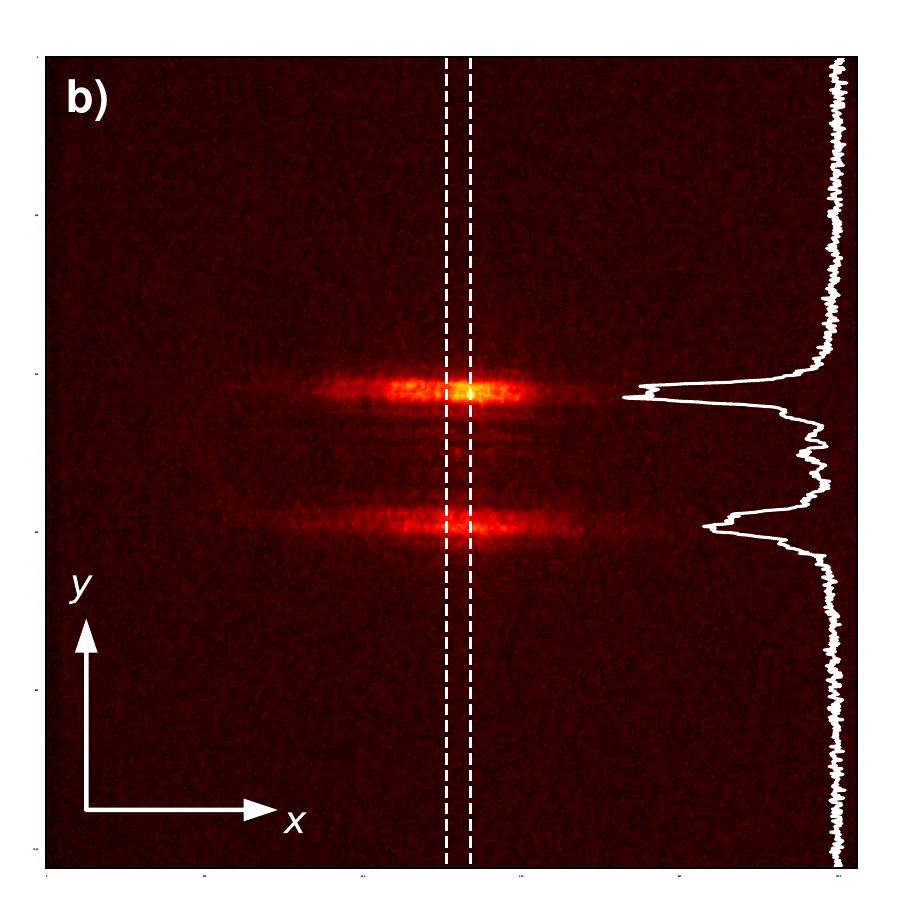}}

\caption{a) Sketch of the schlieren imaging setup. A probe beam (PB) traverses the vapor source (C) cross section and the plasma channel (P) in it through  view ports (W)  along the $z$ direction. A schlieren image is created on the gated camera (GC) by two lenses (L1, L2) in a $4f$ setup and a mask (M) between them. b) Gated camera image with the narrow region of interest that is used to obtain the schlieren signal bounded by vertical dashed lines. The signal $s(y)$ obtained by summing this region along $x$ is plotted on the right of the image.}
\label{fig_schlierensetup}       
\end{figure}

The probe beam passes through two 75 cm focal length lenses arranged in a $4f$ configuration \cite{SalehTeich}, with a circular mask of diameter $D_m = 1.5$~mm placed at the back focal plane of the first lens.
This acts as a high-pass spatial filter, blocking the undeflected probe and transmitting light that has been deflected by the plasma column density gradient.
The mask diameter is also subject to a trade-off: a mask that is too wide reduces signal amplitude and signal-to-noise ratio, while one that is too small may let some undeflected light pass and reduce image contrast.
A gated, image-intensified camera records the resulting schlieren image, triggered 100~ns after the ionizing pulse with a gate duration of 100~ns.
This timing is chosen so that atoms excited but not ionized by the laser return to the ground state by spontaneous decay before the image is taken, while plasma recombination (on a 1--10~$\mu$s timescale at these densities) does not yet alter the plasma density appreciably. A sketch of the geometry is shown in Fig.~\ref{fig_schlierensetup} a).

Since the plasma column is uniform along the ionizing laser propagation direction $x$ on the length scale of the probe beam diameter ($w=3.1$~mm FWHM), the $x$ dependence of the image reflects only the probe beam intensity profile, and the two-dimensional schlieren image can be reduced to a one-dimensional signal $s(y)$ by averaging a narrow horizontal strip of the image along $x$.
This significantly improves the signal-to-noise ratio.
The resulting 1D signal exhibits two prominent peaks at positions $y_1$ and $y_2$ on either side of the plasma column center, arising from the phase gradient at the upper and lower boundaries of the plasma. Light diffracted from the upper and lower edges of the circular mask interferes in the image plane, superimposing a periodic fringe pattern with spatial frequency $f_y \approx D_m / (\lambda_p l)$ (where $l=75$~cm is the distance between the mask and the second lens) on the two-peaked signal structure. This interference distorts the peak shape and is particularly detrimental for narrow plasma channels, where the fringe period can be comparable to the peak separation and the interference amplitude comparable to the peak amplitude itself. A representative camera image and the corresponding signal are shown in Fig.~\ref{fig_schlierensetup} b).

\section{Schlieren image evaluation using symbolic regression}
\label{sec_symreg}

\subsection{Plasma profile models and training database}
\label{sec_profiles}

Our goal is to find analytical formulas relating measurable properties of the schlieren signal to key parameters of the underlying plasma channel, using a training database composed of simulated plasma profile \--- schlieren signal pairs.

To keep the analysis as general as possible, we consider three families of radially symmetric ionization probability profiles $P(r)$ (Eqs.~\ref{eq_profiles}), each parameterized by a core radius $r_0$ and a sheath width parameter ($t_0$, $s_0$ or $l_0$): a Gaussian edge (\ref{eq_profile_gauss}), a sigmoid edge (\ref{eq_profile_sigmoid}), and a cosine edge (\ref{eq_profile_cos}).
\begin{subequations}
\label{eq_profiles}
\begin{equation}
P(r) = P_{max} \cdot \begin{cases}
1, & r \leq r_0, \\[4pt]
\displaystyle\exp\!\left(-\frac{(r-r_0)^2}{t_0^2}\right), & r > r_0,
\end{cases}
\label{eq_profile_gauss}
\end{equation}
\begin{equation}
P(r) = \frac{P_{max}}{1+\exp\!\left(\dfrac{r-r_0}{s_0}\right)},
\label{eq_profile_sigmoid}
\end{equation}
\begin{equation}
P(r) = P_{max}\begin{cases}
1, & r \leq r_0, \\[4pt]
\frac{1}{2}\!\left(1+\cos\!\frac{(r-r_0)\pi}{l_0}\right), & r_0 < r \leq r_0{+}l_0, \\[4pt]
0, & r > r_0{+}l_0.
\end{cases}
\label{eq_profile_cos}
\end{equation}
\end{subequations}
Here $r = \sqrt{(y-y_0)^2+z^2}$, where $y_0$ is the location of the plasma axis in the vertical direction. Numerical simulations \cite{Demeter2019,Demeter2021} and experimental evidence \cite{Adli2019v2} indicate that just after the ionizing pulse (i.e. for delays such that recombination processes can be neglected), the atoms in the plasma channel core are singly ionized with a probability very close to 100\%, so the ionization fraction $P_{max}$ is taken to be 1 in what follows.
The Gaussian edge profile thus reduces to the form used in \cite{Demeter2024}.

Rather than working with the shape-specific parameters $r_0, t_0, s_0, l_0$ directly --- which have different magnitudes across profile types with similar physical extension and shape --- we characterize each plasma sample by two profile-agnostic quantities.
The first is the equivalent radius $R_{eq}$, defined by the area under the radial profile:
\begin{equation}
\label{eq_req}
\pi R_{eq}^2 = 2\pi \int_0^\infty P(r)\,r\,dr.
\end{equation}
The second quantity characterizes the sheath width.
We tested two alternatives: i) the width $W_{91}$, defined as the radial distance between the points where the ionization probability decreases from $P(r) = 0.9$ to $P(r) = 0.1$, and ii) the width defined by the inverse of the maximum radial derivative, $(|dP/dr|_{\max})^{-1}$, which directly measures the sharpness of the plasma boundary that the probe beam reacts to.
Both quantities are proportional to the sheath width parameter for a given profile type; for example, $(|dP/dr|_{\max})^{-1} = t_0\sqrt{e/2}$, $4s_0$ and $2l_0/\pi$ for the Gaussian, sigmoid and cosine edges, respectively.

We generated two training datasets using these profile families, for the two vapor density values ($\mathcal{N} = 5\times10^{14}$ and $7\times10^{14}$~cm$^{-3}$) corresponding to the experimental data in \cite{Demeter2024}. For each sample, the schlieren signal $s(y)$ was computed numerically using standard Fourier optics \cite{SalehTeich}, as described there. The core radius $r_0$ and the sheath width parameter were sampled uniformly over ranges that yield similar coverage of $R_{eq}$ and $W_{91}$ values across all three profiles, while $y_0$ was drawn from a normal distribution with a spread chosen according to the measured variability of the experimental plasma channel position. In parallel, we also generated two equivalent control datasets to be able to verify the performance of various formulas on independent data.  

For some parameter combinations, the two-peaked signal structure described in Sec.~\ref{sec_setup} is absent. This occurs if $y_0$ and/or the plasma radius are large enough that one of the plasma edges falls where the probe beam intensity is already too weak, or if the plasma radius and the sheath width are both so small that the two peaks are not resolved. Because such samples cannot be evaluated by the fitting procedure of Sec.~\ref{sec_fitpars}, such samples are discarded. 

\subsection{Signal fitting and predictor variables}
\label{sec_fitpars}

To apply symbolic regression, each signal must first be condensed into a small number of predictor variables, from which we attempt to infer the plasma channel properties $R_{eq}$ and the sheath width.
This is in contrast to the DNN approach of \cite{Biro2023}, where the full 1024-point raw signal vector served as the network input; a small set of informative predictors is essential for the regression to search the space of candidate formulas efficiently, and it keeps the resulting formulas interpretable.

Most signals in the database display the characteristic structure discussed above: two large peaks with the interference pattern superposed, the modulation due to the interference being most significant for small core radii $r_0$.
The peaks are somewhat asymmetric, and the asymmetry is more significant for sharper plasma edges (narrow sheaths).
These observations suggest fitting the signal with a double-peaked model function that allows different width parameters on the two sides of each maximum.
Among the model functions we tested --- symmetric and asymmetric double Gaussians and double Lorentzians --- the asymmetric double Lorentzian gives the best fit quality to both the simulated and the experimentally observed signals:
\begin{eqnarray}
\label{eq_asymlorentz}
\hat{s}(y) &=& \sum_{k=1,2} \frac{A_k}{1 + \left(\dfrac{y-y_k}{\gamma_k(y)}\right)^2},\nonumber\\
\gamma_k(y) &=& \begin{cases}
\gamma_{k,\ell}, & y < y_k, \\
\gamma_{k,r}, & y \geq y_k,
\end{cases}
\end{eqnarray}
where each peak $k$ has amplitude $A_k$, center $y_k$, and separate half-widths $\gamma_{k,\ell}$ (left side) and $\gamma_{k,r}$ (right side); the peaks are ordered such that $y_1 < y_2$.

\begin{figure}[htb]
\centering
\includegraphics[width=0.95\columnwidth]{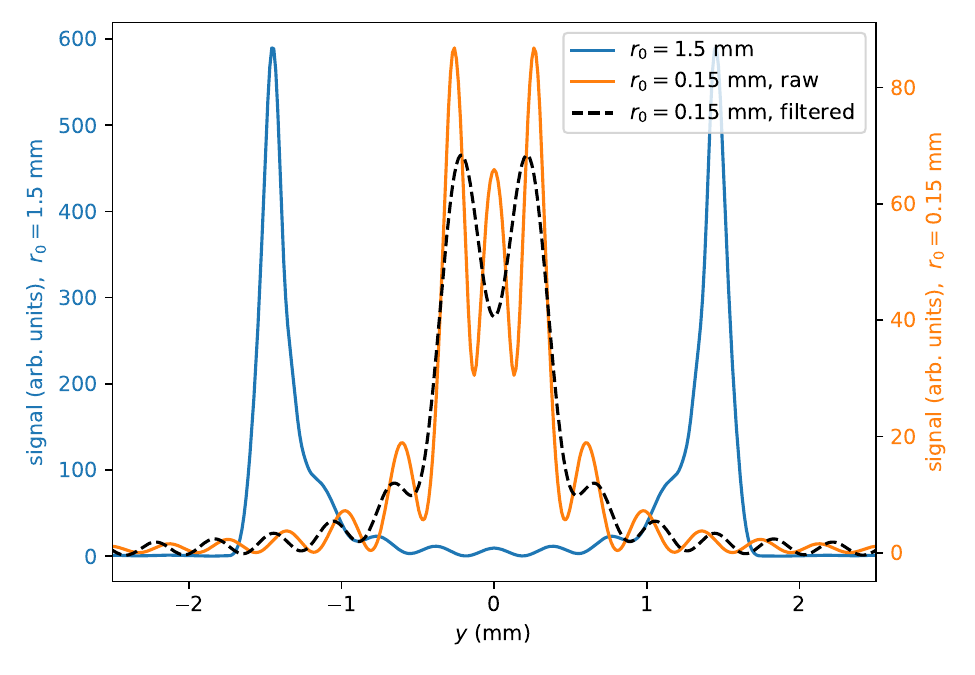}
\caption{Simulated schlieren signals for a Gaussian-edge plasma profile ($y_0=0$, sheath width $t_0=0.15$~mm) with two core radii, illustrating the role of the mask interference and its removal by low-pass filtering. Left axis: $r_0=1.5$~mm (blue), where the two boundary peaks are well separated from the interference fringes riding on the baseline, so filtering is unnecessary. Right axis: $r_0=0.15$~mm (orange), where the boundary peaks lie close together and are strongly modulated by the interference; both the raw signal (solid) and the low-pass filtered signal (black dashed) are shown, illustrating how filtering recovers the underlying double-peak structure in this case. Note that the two signals are plotted on different vertical scales (left/right axis, respectively).}
\label{fig_samplesignals}
\end{figure}

Fitting Eq.~\eqref{eq_asymlorentz} is complicated by the interference: for narrow plasma channels the fit can converge to one of the sharp interference peaks and produce outlier parameters.
The interference can be removed by low-pass filtering the signal below the fringe frequency $D_m/(\lambda_p l)$; this, however, distorts the sharp, high peaks produced by plasma columns with sharp boundaries.
We therefore first assess the importance of the interference for each signal: we compute the filtered signal $\mathcal{F}(s)$ and the ratio of the integrated filtering residual to the filtered signal maximum,
\begin{equation}
\label{eq_ratio}
\rho = \frac{\displaystyle\int |s(y) - \mathcal{F}(s)(y)|\,dy}{\max_y \mathcal{F}(s)(y)},
\end{equation}
which measures the relative weight of the interference content.
When $\rho$ exceeds an empirically chosen threshold, Eq.~\eqref{eq_asymlorentz} is fitted to the filtered signal; otherwise it is fitted to the raw signal.
This hybrid approach balances extracting information from sharp, asymmetric peaks against avoiding false fits due to the interference. Figure \ref{fig_samplesignals} illustrates this point.

\FloatBarrier

The fits reveal a small systematic asymmetry in the simulated signals: the inner widths (the half-widths facing the plasma center, $\gamma_{1,r}$ and $\gamma_{2,\ell}$) tend to be slightly larger than the outer half-widths. From the eight fit parameters we derive the following amplitude-weighted predictor variables:
\begin{eqnarray}
\Delta &=& y_2 - y_1, \label{eq_Delta}\\
W_{in} &=& \frac{A_1 \gamma_{1,r} + A_2 \gamma_{2,\ell}}{A_1 + A_2}, \\
W_{out} &=& \frac{A_1 \gamma_{1,\ell} + A_2 \gamma_{2,r}}{A_1 + A_2}, \\
W &=& \frac{W_{in}+W_{out}}{2}, \\
W_{diff} &=& W_{in} - W_{out}. \label{eq_Wdiff}
\end{eqnarray}
Amplitude weighting mitigates the effect of one peak falling in a region of low probe beam intensity, where its fitted parameters are less reliable.
These five quantities --- $\Delta$, $W_{in}$, $W_{out}$, $W$, $W_{diff}$ --- serve as the inputs to the symbolic regression.
Only three of them are independent, but supplying the derived combinations as well allows the regression to discover simpler expressions.
Note that all five predictor variables are position and width differences and therefore do not depend on the absolute magnitude of the signal peaks; the amplitudes $A_1, A_2$ themselves are not used as predictors. This is a deliberate choice: the overall signal amplitude is set primarily by the probe laser power, which was not monitored continuously during the measurements.

\subsection{Predicting $R_{eq}$}
\label{sec_req}

Symbolic regression was applied to both the $\mathcal{N}=5\cdot 10^{14}\mathrm{~ cm}^{-3}$ and $\mathcal{N}=7\cdot 10^{14}\mathrm{~ cm}^{-3}$ datasets to discover formulas predicting $R_{eq}$ from the predictor variables $\{\Delta, W, W_{in}, W_{out}, W_{diff}\}$.  We used the PySR package \cite{Cranmer2023} to perform the symbolic regression.
PySR allows the user to decide what complexity is attached to various operators, functions and constants in the formulas. Various complexity settings were explored, as well
as the inclusion of various nonlinear functions (power, square root, exponential, log).
Predictor sets of independent quantities $\{ \Delta, W, W_{diff} \}$ and $\{ \Delta, W_{in}, W_{out} \}$ were tried, as well as the full set $\{ \Delta, W, W_{in}, W_{out}, W_{diff} \}$.
This is useful in our case, because PySR may attach different complexities to the same formula if derived from different sets, whereas for a human observer they may be of equivalent complexity even in mixed form. Because of the non-deterministic evolutionary algorithm, it is also useful to run the search several times, as different formulas may (and will) eventually emerge from the search. As a loss function we used the mean absolute error, denoted $\sigma$ throughout, the same measure used in \cite{Biro2023}.

\begin{figure}[htb]
\includegraphics[width=\columnwidth]{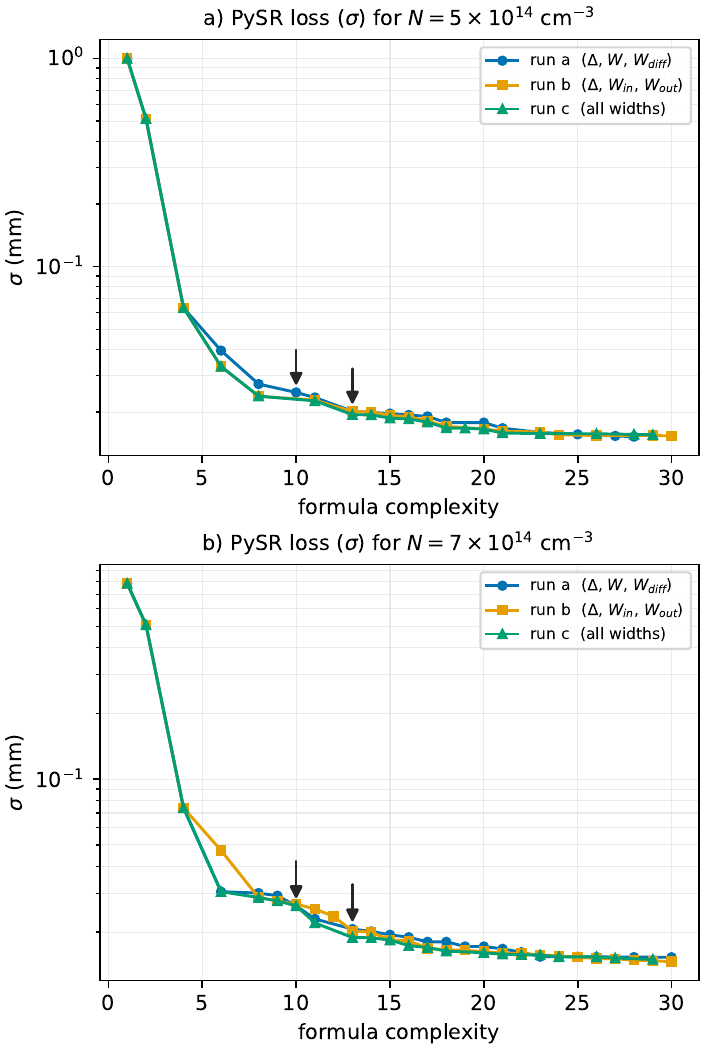}
\caption{Mean absolute error loss vs. formula complexity during regression runs by PySR for $R_{eq}$ using different width parameter predictors, for vapor density a) $\mathcal{N}=5\times10^{14}$ cm$^{-3}$ and  b) $\mathcal{N}=7\times10^{14}$ cm$^{-3}$. Black arrows mark the location of Eqs. \ref{Req_formulas}.}
\label{fig_Reqfits_loss}       
\end{figure}

So which of the formulas returned should we use as the final ``predictor'' for $R_{eq}$? There are several considerations. First, we may plot the decrease of the loss on the dataset used for fitting (Fig. \ref{fig_Reqfits_loss}). The plot has a long tail where the loss decreases with complexity only marginally, so it seems obvious to select formulas that are just after the last major decrease of the loss function, that being in some way an optimum with respect to loss and complexity together. (The different PySR runs shown in Fig.~\ref{fig_Reqfits_loss} used three versions of the width parameters as predictor variables besides $\Delta$.)

It is also instructive to compare the formulas found across the various runs. Some formulas may surface in several runs --- formulas that can be transformed into the same form are treated as identical of course. This may happen when the $W_{in}, W_{out} \leftrightarrow W, W_{diff}$ predictors yield formulas that reduce to the same expression and such recurring formulas are preferred candidates.
Especially so, if they appear in runs for both vapor densities (the fitting constants may of course differ between the two cases) as this means that the functional relationship is applicable in a broad parameter range. 
(Formulas with complexity as high as 18 appear in runs for both densities for the $R_{eq}$ case.)
In the absence of the same formula appearing in different density runs,
we may test the cross-transfer of formulas found for the two densities: if a formula found for $\mathcal{N}=5$ performs similarly well, with recalculated constants, on the $\mathcal{N}=7$ control data (or vice versa), it should be preferred over formulas that do not transfer well. The transferred formula may perform slightly worse than the ``native'' formula for that density and complexity, but typically not by much; if a formula transfers well, it is reasonable to assume that, for any density, re-optimizing its constants on a modest new database is sufficient to obtain a good predictor.

Another consideration is interpretability. Even if we do not discover a formula which excels universally compared to all other variants of similar complexity, it may happen that some structure in some of the formulas can be given clear physical meaning.

Out of the formulas returned by PySR, we here select just two: one that is purely linear in the predictor variables and contains a single adjustable constant, Eq.~\ref{Req_formula1}, and one that contains two adjustable constants and is slightly nonlinear, Eq.~\ref{Req_formula2}. 
\begin{subequations}
\label{Req_formulas}
\begin{equation}
f_1 = C_1 \left(2\Delta + W\right) - W_{diff}, \label{Req_formula1}
\end{equation}
\begin{equation}
f_2 = (C_2 + C_1  \Delta)\cdot(\Delta+W) - W_{diff}. \label{Req_formula2}
\end{equation}
\end{subequations}

\begin{figure}[htb]
\includegraphics[width=\columnwidth]{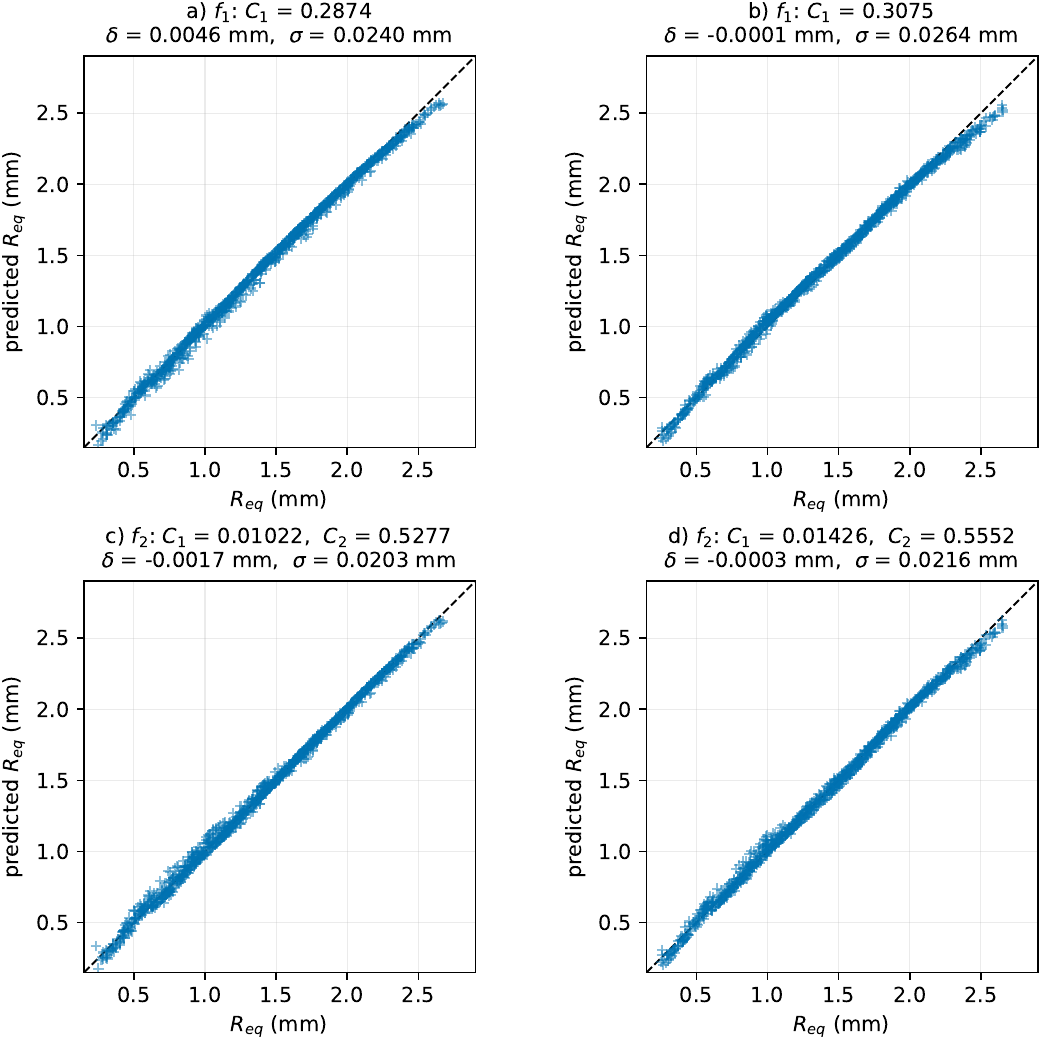}
\caption{Plots of predicted vs. real values of $R_{eq}$ on the control datasets. 
a) and b): The predictive performance of the linear formula Eq.~\ref{Req_formula1} for density a) $\mathcal{N}=5\times10^{14}$ cm$^{-3}$ and  b) $\mathcal{N}=7\times10^{14}$ cm$^{-3}$. c) and d): The predictive 
performance of the nonlinear formula Eq.~\ref{Req_formula2} for the two densities. 
Dashed black line marks predicted = real $R_{eq}$. }
\label{fig_Reqfits}       
\end{figure}

From the numerical values of the constants (displayed in Figs. \ref{fig_Reqfits}) it is clear that the dominant factor in determining the plasma radius is the peak distance $\Delta$, the prediction then being adjusted by the peak width parameters. The inner and outer peak width values do not contribute equally, confirming that the asymmetry of the peaks carries important information. 
The constant $C_1$ in the second formula is small, so the nonlinearity acts as a minor correction to the dominant linear terms; the linear form obtained by setting $C_1=0$ is not identical to the first formula, though it too appears in the Pareto front returned by PySR for some runs. Both formulas recur multiple times in the results and appear for both densities with comparable loss values, i.e. the same functional form (with slightly different values of the constants) yields very similar overall loss for predicting $R_{eq}$ (see Table~\ref{tab_Req} and Fig.~\ref{fig_Reqfits}). Note that slightly different complexities may be attached to the same formula in different runs, depending on which set of width variables was supplied as predictors.

\FloatBarrier

It is clear from Fig.~\ref{fig_Reqfits} that $R_{eq}$ can be predicted with very good accuracy for both densities by simple formulas using just one or two fit parameters.
The mean absolute errors in Table~\ref{tab_Req} should be compared with the pixel size of the camera, 0.013~mm, which sets the spatial resolution of the recorded schlieren signal: both formulas predict $R_{eq}$ to within about twice the pixel size. The slight nonlinear correction in Eq.~\ref{Req_formula2} is also well justified by the plots (compare Fig.~\ref{fig_Reqfits}a,b for the linear formula to c,d for the nonlinear one): it corrects slightly larger deviations towards the high and low ends of the $R_{eq}$ range.

\begin{table}[htb]
\centering
\caption{Prediction accuracy of $f_1$ (Eq.~\ref{Req_formula1}) and $f_2$ (Eq.~\ref{Req_formula2}) for $R_{eq}$, evaluated on the control datasets for both vapor densities (constants optimized separately for each density). $\sigma$ is the mean absolute error and $\delta$ the signed mean error, both in mm; see Fig.~\ref{fig_Reqfits}.}
\label{tab_Req}
\begin{tabular}{|c|cc|cc|}
\hline
& \multicolumn{2}{c|}{$\mathcal{N}=5$} & \multicolumn{2}{c}{$\mathcal{N}=7$} \\
formula &  $\sigma$ & $\delta$ & $\sigma$ & $\delta$ \\
\hline
$f_1$ &  0.0240 & 0.0046  & 0.0264 & $-$0.0001 \\
$f_2$ &  0.0203 & $-$0.0017 & 0.0216 & $-$0.0003 \\
\hline
\end{tabular}
\end{table}
\FloatBarrier

\subsection{Predicting the sheath width}
\label{sec_width}

\begin{figure}[htb]
\includegraphics[width=\columnwidth]{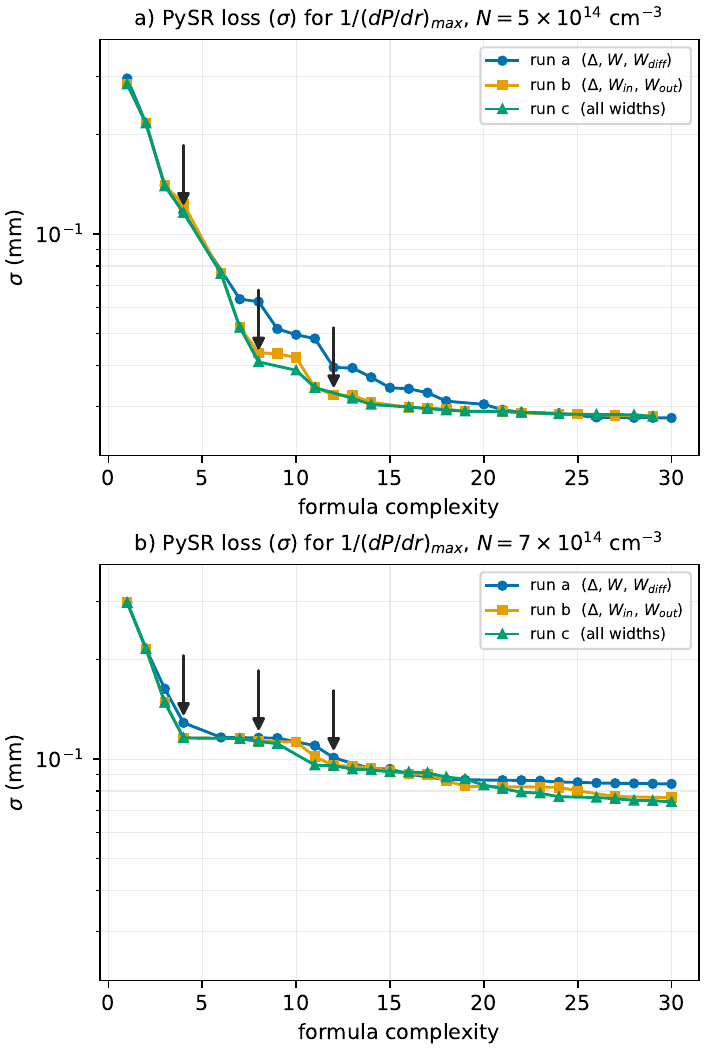}
\caption{Mean absolute error loss vs. formula complexity during regression runs by PySR for $(dP/dr)^{-1}$ using different width parameter predictors, for vapor density a) $\mathcal{N}=5\times10^{14}$ cm$^{-3}$ and  b) $\mathcal{N}=7\times10^{14}$ cm$^{-3}$. Black arrows mark the location of Eqs. \ref{dPdr_formulas}.}
\label{fig_dPdrfits_loss}       
\end{figure}

Predicting the sheath transition width proves harder than predicting $R_{eq}$. From the two candidates $(dP/dr)^{-1}$ proved to be more predictable. When searching for formulas attempting to predict $W_{91}$ and separating the performance with respect to the three plasma profile functions, we repeatedly found large performance gaps.  
This was not the case for $(dP/dr)^{-1}$, the quantity we use exclusively in what follows. 
Symbolic regression over the same dataset returns multiple formulas of similar complexity and prediction accuracy for $(dP/dr)^{-1}$, but with very different functional forms. Plots of the loss functions shown in Figs.~\ref{fig_dPdrfits_loss} a) and b) show that, while for $\mathcal{N}=5$ the loss converges to $\sigma<0.03$, the $\mathcal{N}=7$ fits eventually plateau at roughly three times this value, confirming that the sheath width is inherently harder to predict from these predictor variables at the higher density. Three formulas were selected for comparison from these runs:

\begin{subequations}
\label{dPdr_formulas}
\begin{equation}
g_1 = C_1 W \label{dPdr_formula1}
\end{equation}
\begin{equation}
g_2 = C_1 W_{out}\Delta+W \label{dPdr_formula2}
\end{equation}
\begin{equation}
g_3 = C_1 W_{out}\Delta\cdot\exp(-C_2\Delta) \label{dPdr_formula3}
\end{equation}
\end{subequations}

\begin{figure}[htb]
\includegraphics[width=\columnwidth]{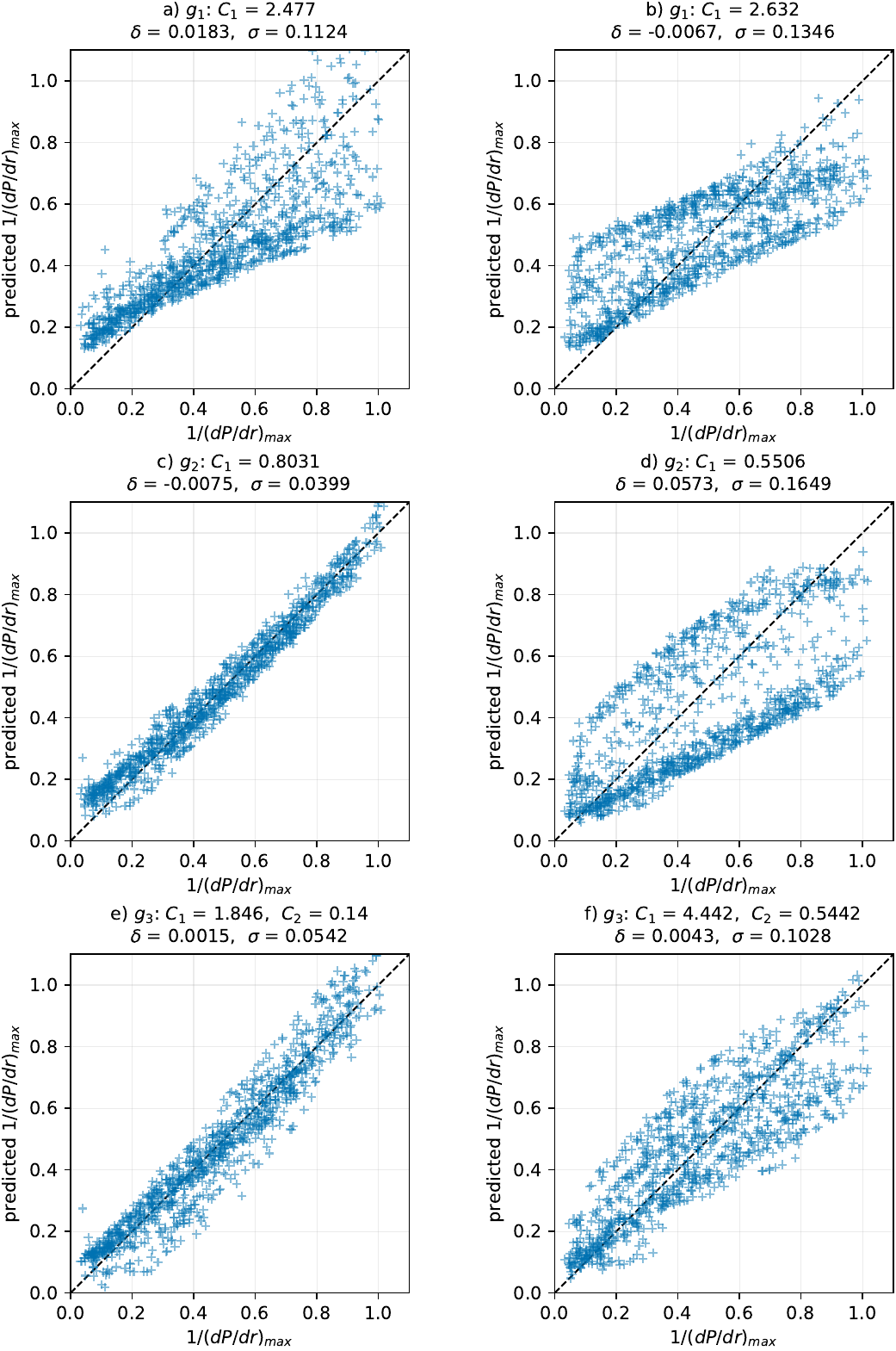}
\caption{Plots of predicted vs. real values of $1/|dP/dr|_{max}$ on the control datasets. 
a) and b): The predictive performance of Eq.~\ref{dPdr_formula1} for density a) $\mathcal{N}=5\times10^{14}$ cm$^{-3}$ and  b) $\mathcal{N}=7\times10^{14}$ cm$^{-3}$. c) and d): The predictive 
performance of Eq.~\ref{dPdr_formula2} for the two densities. e) and f): The predictive 
performance of Eq.~\ref{dPdr_formula3} for the two densities.
Dashed black line marks predicted = real $1/|dP/dr|_{max}$. 
}
\label{fig_dPdrfits}       
\end{figure}

Their prediction accuracy is summarized in Table~\ref{tab_dPdr} and illustrated in Fig.~\ref{fig_dPdrfits}.
Eq.~\ref{dPdr_formula1} is the simplest of the three and the least accurate: it predicts the sheath width from the overall peak width alone, yet it is the only formula that turns up in both the $\mathcal{N}=5$ and $\mathcal{N}=7$ regression runs. Eq.~\ref{dPdr_formula2} appears multiple times in the $\mathcal{N}=5$ regression runs and is still very simple. It works well for the $\mathcal{N}=5$ density, with $\sigma$ close to the limit of predictability in that case; transferred to $\mathcal{N}=7$, however, it fails, with $\sigma$ higher than even that produced by Eqs.~\ref{dPdr_formula1}. Finally, Eq.~\ref{dPdr_formula3}, found for the $\mathcal{N}=7$ density, is fairly simple and works close to the predictability limit for that density; when transferred to $\mathcal{N}=5$ it performs substantially worse than Eq.~\ref{dPdr_formula2}, though still much better than it does for the $\mathcal{N}=7$ density for which it was originally derived. This underlines that, in this case, predictability is inherently worse for the $\mathcal{N}=7$ density.

The loss of predictability is fundamentally tied to the refractive index, i.e. the optical density, through the phase difference accumulated by the probe beam. The refractive index depends strongly on the  
vapor density because the optical linewidth broadens due to collisional broadening.  By tuning the probe wavelength slightly farther from resonance, the refractive index at the higher density can be matched to that at the lower density. Indeed, we find that for a probe wavelength of $\lambda=780.336$~nm --- slightly larger than the value used for both densities in the original experiment \cite{Demeter2024} --- the $\mathcal{N}=7$ signals can be evaluated directly with the formulas found for the $\mathcal{N}=5$ regression runs, and both plasma parameters are predicted with only marginally lower accuracy. 

\FloatBarrier

\FloatBarrier

\begin{table}[htb]
\centering
\caption{Prediction accuracy of $g_1$ (Eq.~\ref{dPdr_formula1}), $g_2$ (Eq.~\ref{dPdr_formula2}) and $g_3$ (Eq.~\ref{dPdr_formula3}) for $(dP/dr)^{-1}_{max}$, evaluated on the control datasets for both vapor densities (constants re-optimized for each density). $\sigma$ is the mean absolute error and $\delta$ the signed mean error; see Fig.~\ref{fig_dPdrfits}.}
\label{tab_dPdr}
\begin{tabular}{|c|cc|cc|}
\hline
&  \multicolumn{2}{c|}{$\mathcal{N}=5$} & \multicolumn{2}{c}{$\mathcal{N}=7$} \\
formula  & $\sigma$ & $\delta$ & $\sigma$ & $\delta$ \\
\hline
$g_1$   & 0.1124 & 0.0183  & 0.1346 & $-$0.0067 \\
$g_2$   & 0.0399 & $-$0.0075 & 0.1649 & 0.0573 \\
$g_3$  & 0.0542 & 0.0015  & 0.1028 & 0.0043 \\
\hline
\end{tabular}
\end{table}
\FloatBarrier

\subsection{Comparison with previous approaches}

The symbolic regression approach offers several practical advantages over the methods used previously for
evaluating schlieren imaging measurements \cite{Biro2023,Demeter2024}.
The DNN approach of \cite{Biro2023} requires a very large training database ($\sim 10^6$ samples) and a complex neural network with $10^6$--$10^7$ trainable parameters (weights). The required numerical effort is sizeable: it necessitates several days of processing time on modern, high-power GPU clusters. No human input is required in deriving predictor variables from the signals, since the input to the DNN is the signal vector itself, but one has to experiment with various network architectures and sizes, each with its own training run, and add noise to the samples to make the predictions robust to experimental noise. The resulting trained networks predict the plasma parameters very accurately, but a full re-training run is required whenever the experimental parameters change by more than a few percent. (It must be noted that the DNN approach was tested only for a single plasma density, $\mathcal{N}=5\times10^{14}$~cm$^{-3}$, and only for the less general, single Gaussian sheath profile (Eq.~\ref{eq_profile_gauss}).) Furthermore, in \cite{Biro2023} plasma parameters were predicted simultaneously by the same DNN, so to obtain comparable relative accuracy the parameters had to be scaled to roughly the same magnitude.

In contrast, the current symbolic regression procedure requires a training database of just $10^3$ samples; sample generation takes about two minutes and the regression itself about 20 minutes on an ordinary desktop computer. Once functional forms have been derived for a range of densities (or other varying experimental parameters), adapting to new conditions requires only a quick re-fit of 1--3 constants on a modest new database. The resulting formulas are also fully explicit, which facilitates cross-checking and error propagation. On the other hand, human input (and experimentation) is needed in determining the few parameters of the signal that are used for plasma parameter prediction.

In \cite{Demeter2024}, a Gaussian plasma profile was assumed, symmetric peak parameters were derived from the signals, and it was observed that, to a fairly good approximation:
\begin{eqnarray}
 \Delta &= M_{12}r_0 t_0 + M_1 r_0 + M_2 t_0 + B\nonumber\\
 W &= Q_{12} r_0 t_0 + Q_1 r_0 + Q_2 t_0 + P
 \label{eq_schlierenconsts}
\end{eqnarray}
Here $\Delta$ is identical to the one used in the present work, and $W$ is an amplitude-weighted mean peak width derived from the filtered signal without curve fitting. The approach was then to use the fitting database to fit the 8 free parameters of the model (separately for both plasma densities) and invert the relationships to find the plasma parameters $r_0, t_0$. This approach is numerically light, just like symbolic regression, but it uses 8 fitting constants, and is more restrictive in that it assumes a single plasma profile. Furthermore, it required considerable experimentation and human work to find the approximate relationship Eq. \ref{eq_schlierenconsts} which still does not exploit the information obtainable from the asymmetry of the peaks.

The final test for the method is using it to evaluate real experimental data and comparing the results with values expected from numerical simulations.  
Long range propagation calculations of resonant, ionizing laser pulses in Rb vapor suggest \cite{Demeter_tbp} that
after some distance, the propagating pulse generates a plasma channel whose equivalent radius $R_{eq}$ oscillates around the value
\begin{equation}
\label{eq_powerlaw}
 R_{eq}(z) = r_{ref}\cdot \left(\frac{E(z)}{E_{ref}}\right)^C,
\end{equation}
where $E(z)$ is the energy contained in the pulse at propagation distance $z$, $E_{ref}$ is an arbitrary reference energy  and $r_{ref}$ is the plasma radius at this reference energy value. We conveniently choose the reference energy to be $E_{ref}=1$~mJ and write the power-law relationship as $R_{eq} = r_1\cdot E^C$. This prediction can be tested by measuring the plasma radius using schlieren imaging and measuring the energy of the propagating pulse simultaneously. Suitable experimental data is available: the schlieren measurements of \cite{Demeter2024} at vapor density $\mathcal{N}=5\times10^{14}$~cm$^{-3}$, performed very close to the end of the vapor source at AWAKE, included a simultaneous measurement of the transmitted pulse energy.

\begin{figure}[htb]
\includegraphics[width=\columnwidth]{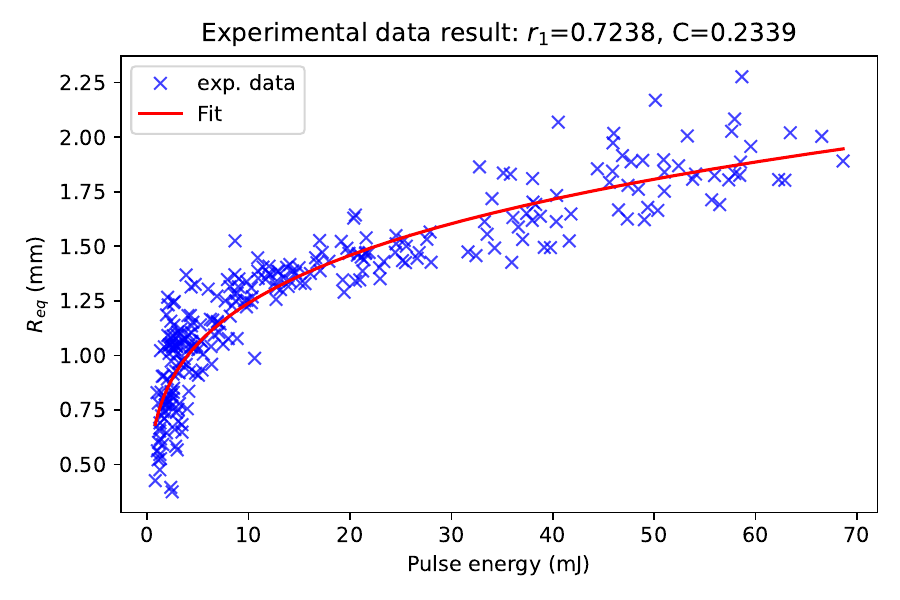}
\caption{Equivalent plasma channel radius $R_{eq}$, obtained from the experimental schlieren signals of \cite{Demeter2024} using Eq.~\ref{Req_formula2}, as a function of the simultaneously measured transmitted pulse energy (crosses). The solid line is the fitted power law $R_{eq} = r_1 E^C$ with $r_1 = 0.724$~mm and $C=0.234$.}
\label{fig_powerlaw_fit}   
\end{figure}

We used the present method that obtains $R_{eq}$ directly from the schlieren data to re-evaluate these measurements and obtain experimental values for $r_1$ and $C$ for a comparison with theory. Figure \ref{fig_powerlaw_fit} shows the experimental data after processing with formula \ref{Req_formula2} (with the constants optimized for this density) and the power-law fit to the datapoints, which yields $r_1 = 0.724 \mathrm{~mm}$, $C=0.234$. These values lie close to the values obtained by similar fits to simulation data: $r_1 = 0.696 \mathrm{~mm}, C=0.2125$.

\section{Summary}
\label{sec_summary}

We have presented a symbolic regression approach for evaluating near-resonant schlieren images of the laser-ionized plasma channel of the AWAKE experiment.
The plasma column is characterized by two profile-agnostic parameters --- the equivalent radius $R_{eq}$ and the sheath width $(|dP/dr|_{max})^{-1}$ --- and databases of simulated schlieren signals are generated using three different families of plasma edge profiles.
Each signal is condensed into five predictor variables by fitting an asymmetric double Lorentzian to the two-peaked signal structure, with low-pass filtering applied when the interference caused by the mask dominates the signal.
Symbolic regression with the PySR package then yields compact analytical formulas connecting the predictor variables to the plasma parameters.

For the equivalent radius, a purely linear formula with a single adjustable constant predicts $R_{eq}$ with a mean absolute error of about twice the camera pixel size, and a slightly nonlinear formula with two constants performs even better.
Both forms recur across regression runs for the two vapor densities studied, with constants re-optimized for each density.
The sheath width is inherently harder to predict, especially at the higher vapor density, where enhanced collisional broadening of the atomic line degrades predictability; re-tuning the probe wavelength slightly farther from resonance restores the accuracy and even allows the direct transfer of the formulas found for the lower density.
Compared with the deep neural network approach of \cite{Biro2023}, the method requires a training database smaller by three orders of magnitude and runs on an ordinary desktop computer in well under an hour, while yielding fully explicit, interpretable formulas; compared with the bilinear model of \cite{Demeter2024}, it needs fewer fitting constants and does not assume a single specific plasma edge profile.
Finally, we applied the method to experimental data, re-evaluating schlieren measurements taken near the end of the AWAKE vapor source. Plasma radius values obtained with the present method corroborate 
the theoretically predicted power-law relationship between $R_{eq}$ and the ionizing pulse energy, with fitted constants close to those obtained from simulations.

\section*{Acknowledgment}
On behalf of Project Awakelaser we are grateful for the usage of HUN-REN Cloud \cite{Heder2022} which helped us achieve the results published in this paper.

\section*{Declaration of generative AI and AI-assisted technologies in the manuscript preparation process}

 During the preparation of this work the author used Claude Code in order to generate Figure 1 a), as well as to improve the structure and language of some paragraphs. After using this tool/service, the author reviewed and edited the content as needed and takes full responsibility for the content of the published article.

\clearpage

\bibliography{/home/gdemeter/fiz/manuscript/bibliography_pulseprop}

\begin{thebibliography}{25}%
\makeatletter
\providecommand \@ifxundefined [1]{%
 \@ifx{#1\undefined}
}%
\providecommand \@ifnum [1]{%
 \ifnum #1\expandafter \@firstoftwo
 \else \expandafter \@secondoftwo
 \fi
}%
\providecommand \@ifx [1]{%
 \ifx #1\expandafter \@firstoftwo
 \else \expandafter \@secondoftwo
 \fi
}%
\providecommand \natexlab [1]{#1}%
\providecommand \enquote  [1]{``#1''}%
\providecommand \bibnamefont  [1]{#1}%
\providecommand \bibfnamefont [1]{#1}%
\providecommand \citenamefont [1]{#1}%
\providecommand \href@noop [0]{\@secondoftwo}%
\providecommand \href [0]{\begingroup \@sanitize@url \@href}%
\providecommand \@href[1]{\@@startlink{#1}\@@href}%
\providecommand \@@href[1]{\endgroup#1\@@endlink}%
\providecommand \@sanitize@url [0]{\catcode `\\12\catcode `\$12\catcode
  `\&12\catcode `\#12\catcode `\^12\catcode `\_12\catcode `\%12\relax}%
\providecommand \@@startlink[1]{}%
\providecommand \@@endlink[0]{}%
\providecommand \url  [0]{\begingroup\@sanitize@url \@url }%
\providecommand \@url [1]{\endgroup\@href {#1}{\urlprefix }}%
\providecommand \urlprefix  [0]{URL }%
\providecommand \Eprint [0]{\href }%
\providecommand \doibase [0]{http://dx.doi.org/}%
\providecommand \selectlanguage [0]{\@gobble}%
\providecommand \bibinfo  [0]{\@secondoftwo}%
\providecommand \bibfield  [0]{\@secondoftwo}%
\providecommand \translation [1]{[#1]}%
\providecommand \BibitemOpen [0]{}%
\providecommand \bibitemStop [0]{}%
\providecommand \bibitemNoStop [0]{.\EOS\space}%
\providecommand \EOS [0]{\spacefactor3000\relax}%
\providecommand \BibitemShut  [1]{\csname bibitem#1\endcsname}%
\let\auto@bib@innerbib\@empty
\bibitem [{\citenamefont {Settles}(2001)}]{Schlierenbook}%
  \BibitemOpen
  \bibfield  {author} {\bibinfo {author} {\bibfnamefont {G.~S.}\ \bibnamefont
  {Settles}},\ }\href@noop {} {\emph {\bibinfo {title} {Schlieren and
  Shadowgraph Techniques - Visualizing Phenomena In Transparent Media}}}\
  (\bibinfo  {publisher} {Springer-Verlag, Berlin, Heidelberg},\ \bibinfo
  {year} {2001})\BibitemShut {NoStop}%
\bibitem [{\citenamefont {Iwase}\ \emph {et~al.}(1998)\citenamefont {Iwase},
  \citenamefont {Süß}, \citenamefont {Hoffmann}, \citenamefont {Roth},
  \citenamefont {Stöckl}, \citenamefont {Geissel}, \citenamefont {Seelig},\
  and\ \citenamefont {Bock}}]{Iwase1998}%
  \BibitemOpen
  \bibfield  {author} {\bibinfo {author} {\bibfnamefont {O.}~\bibnamefont
  {Iwase}}, \bibinfo {author} {\bibfnamefont {W.}~\bibnamefont {Süß}},
  \bibinfo {author} {\bibfnamefont {D.~H.~H.}\ \bibnamefont {Hoffmann}},
  \bibinfo {author} {\bibfnamefont {M.}~\bibnamefont {Roth}}, \bibinfo {author}
  {\bibfnamefont {C.}~\bibnamefont {Stöckl}}, \bibinfo {author} {\bibfnamefont
  {M.}~\bibnamefont {Geissel}}, \bibinfo {author} {\bibfnamefont
  {W.}~\bibnamefont {Seelig}}, \ and\ \bibinfo {author} {\bibfnamefont
  {R.}~\bibnamefont {Bock}},\ }\href {\doibase 10.1088/0031-8949/58/6/017}
  {\bibfield  {journal} {\bibinfo  {journal} {Physica Scripta}\ }\textbf
  {\bibinfo {volume} {58}},\ \bibinfo {pages} {634} (\bibinfo {year}
  {1998})}\BibitemShut {NoStop}%
\bibitem [{\citenamefont {Clayton}\ \emph {et~al.}(1998)\citenamefont
  {Clayton}, \citenamefont {Tzeng}, \citenamefont {Gordon}, \citenamefont
  {Muggli}, \citenamefont {Mori}, \citenamefont {Joshi}, \citenamefont {Malka},
  \citenamefont {Najmudin}, \citenamefont {Modena}, \citenamefont {Neely},\
  and\ \citenamefont {Dangor}}]{Clayton1998}%
  \BibitemOpen
  \bibfield  {author} {\bibinfo {author} {\bibfnamefont {C.~E.}\ \bibnamefont
  {Clayton}}, \bibinfo {author} {\bibfnamefont {K.-C.}\ \bibnamefont {Tzeng}},
  \bibinfo {author} {\bibfnamefont {D.}~\bibnamefont {Gordon}}, \bibinfo
  {author} {\bibfnamefont {P.}~\bibnamefont {Muggli}}, \bibinfo {author}
  {\bibfnamefont {W.~B.}\ \bibnamefont {Mori}}, \bibinfo {author}
  {\bibfnamefont {C.}~\bibnamefont {Joshi}}, \bibinfo {author} {\bibfnamefont
  {V.}~\bibnamefont {Malka}}, \bibinfo {author} {\bibfnamefont
  {Z.}~\bibnamefont {Najmudin}}, \bibinfo {author} {\bibfnamefont
  {A.}~\bibnamefont {Modena}}, \bibinfo {author} {\bibfnamefont
  {D.}~\bibnamefont {Neely}}, \ and\ \bibinfo {author} {\bibfnamefont {A.~E.}\
  \bibnamefont {Dangor}},\ }\href {\doibase 10.1103/PhysRevLett.81.100}
  {\bibfield  {journal} {\bibinfo  {journal} {Phys. Rev. Lett.}\ }\textbf
  {\bibinfo {volume} {81}},\ \bibinfo {pages} {100} (\bibinfo {year}
  {1998})}\BibitemShut {NoStop}%
\bibitem [{\citenamefont {Honda}\ \emph {et~al.}(2000)\citenamefont {Honda},
  \citenamefont {Katsura}, \citenamefont {Takahashi},\ and\ \citenamefont
  {Kondo}}]{Honda2000}%
  \BibitemOpen
  \bibfield  {author} {\bibinfo {author} {\bibfnamefont {H.}~\bibnamefont
  {Honda}}, \bibinfo {author} {\bibfnamefont {K.}~\bibnamefont {Katsura}},
  \bibinfo {author} {\bibfnamefont {E.}~\bibnamefont {Takahashi}}, \ and\
  \bibinfo {author} {\bibfnamefont {K.}~\bibnamefont {Kondo}},\ }\href@noop {}
  {\bibfield  {journal} {\bibinfo  {journal} {Applied Physics B}\ }\textbf
  {\bibinfo {volume} {70}},\ \bibinfo {pages} {395} (\bibinfo {year}
  {2000})}\BibitemShut {NoStop}%
\bibitem [{\citenamefont {Veloso}\ \emph {et~al.}(2006)\citenamefont {Veloso},
  \citenamefont {Chuaqui}, \citenamefont {Aliaga-Rossel}, \citenamefont
  {Favre}, \citenamefont {Mitchell},\ and\ \citenamefont
  {Wyndham}}]{Veloso2006}%
  \BibitemOpen
  \bibfield  {author} {\bibinfo {author} {\bibfnamefont {F.}~\bibnamefont
  {Veloso}}, \bibinfo {author} {\bibfnamefont {H.}~\bibnamefont {Chuaqui}},
  \bibinfo {author} {\bibfnamefont {R.}~\bibnamefont {Aliaga-Rossel}}, \bibinfo
  {author} {\bibfnamefont {M.}~\bibnamefont {Favre}}, \bibinfo {author}
  {\bibfnamefont {I.~H.}\ \bibnamefont {Mitchell}}, \ and\ \bibinfo {author}
  {\bibfnamefont {E.}~\bibnamefont {Wyndham}},\ }\href {\doibase
  10.1063/1.2206773} {\bibfield  {journal} {\bibinfo  {journal} {Review of
  Scientific Instruments}\ }\textbf {\bibinfo {volume} {77}},\ \bibinfo {pages}
  {063506} (\bibinfo {year} {2006})}\BibitemShut {NoStop}%
\bibitem [{\citenamefont {Batani}\ \emph {et~al.}(2019)\citenamefont {Batani},
  \citenamefont {Santos}, \citenamefont {Forestier-Colleoni}, \citenamefont
  {Mancelli}, \citenamefont {Ehret}, \citenamefont {Trela}, \citenamefont
  {Morace}, \citenamefont {Jakubowska}, \citenamefont {Antonelli},
  \citenamefont {del Sorbo}, \citenamefont {Manclossi},\ and\ \citenamefont
  {Veltcheva}}]{Batani2019}%
  \BibitemOpen
  \bibfield  {author} {\bibinfo {author} {\bibfnamefont {D.}~\bibnamefont
  {Batani}}, \bibinfo {author} {\bibfnamefont {J.}~\bibnamefont {Santos}},
  \bibinfo {author} {\bibfnamefont {P.}~\bibnamefont {Forestier-Colleoni}},
  \bibinfo {author} {\bibfnamefont {D.}~\bibnamefont {Mancelli}}, \bibinfo
  {author} {\bibfnamefont {M.}~\bibnamefont {Ehret}}, \bibinfo {author}
  {\bibfnamefont {J.}~\bibnamefont {Trela}}, \bibinfo {author} {\bibfnamefont
  {A.}~\bibnamefont {Morace}}, \bibinfo {author} {\bibfnamefont
  {K.}~\bibnamefont {Jakubowska}}, \bibinfo {author} {\bibfnamefont
  {L.}~\bibnamefont {Antonelli}}, \bibinfo {author} {\bibfnamefont
  {D.}~\bibnamefont {del Sorbo}}, \bibinfo {author} {\bibfnamefont
  {M.}~\bibnamefont {Manclossi}}, \ and\ \bibinfo {author} {\bibfnamefont
  {M.}~\bibnamefont {Veltcheva}},\ }\href@noop {} {\bibfield  {journal}
  {\bibinfo  {journal} {Journal of Fusion Energy}\ }\textbf {\bibinfo {volume}
  {38}},\ \bibinfo {pages} {299} (\bibinfo {year} {2019})}\BibitemShut
  {NoStop}%
\bibitem [{\citenamefont {Bachmann}\ \emph {et~al.}(2018)\citenamefont
  {Bachmann}, \citenamefont {Martyanov}, \citenamefont {Moody}, \citenamefont
  {Pukhov},\ and\ \citenamefont {Muggli}}]{Bachmann2018}%
  \BibitemOpen
  \bibfield  {author} {\bibinfo {author} {\bibfnamefont {A.-M.}\ \bibnamefont
  {Bachmann}}, \bibinfo {author} {\bibfnamefont {M.}~\bibnamefont {Martyanov}},
  \bibinfo {author} {\bibfnamefont {J.}~\bibnamefont {Moody}}, \bibinfo
  {author} {\bibfnamefont {A.}~\bibnamefont {Pukhov}}, \ and\ \bibinfo {author}
  {\bibfnamefont {P.}~\bibnamefont {Muggli}},\ }\href {\doibase
  10.1016/j.nima.2017.12.062} {\bibfield  {journal} {\bibinfo  {journal}
  {Nuclear Instruments and Methods in Physics Research Section A: Accelerators,
  Spectrometers, Detectors and Associated Equipment}\ }\textbf {\bibinfo
  {volume} {909}},\ \bibinfo {pages} {387–390} (\bibinfo {year} {2018})},\
  \bibinfo {note} {3rd European Advanced Accelerator Concepts workshop
  (EAAC2017)}\BibitemShut {NoStop}%
\bibitem [{\citenamefont {Demeter}\ \emph {et~al.}(2024)\citenamefont
  {Demeter}, \citenamefont {Moody}, \citenamefont {Kedves}, \citenamefont
  {Batsch}, \citenamefont {Bergamaschi}, \citenamefont {Fedosseev},
  \citenamefont {Granados}, \citenamefont {Muggli}, \citenamefont {Panuganti},\
  and\ \citenamefont {{Zevi Della Porta}}}]{Demeter2024}%
  \BibitemOpen
  \bibfield  {author} {\bibinfo {author} {\bibfnamefont {G.}~\bibnamefont
  {Demeter}}, \bibinfo {author} {\bibfnamefont {J.}~\bibnamefont {Moody}},
  \bibinfo {author} {\bibfnamefont {M.}~\bibnamefont {Kedves}}, \bibinfo
  {author} {\bibfnamefont {F.}~\bibnamefont {Batsch}}, \bibinfo {author}
  {\bibfnamefont {M.}~\bibnamefont {Bergamaschi}}, \bibinfo {author}
  {\bibfnamefont {V.}~\bibnamefont {Fedosseev}}, \bibinfo {author}
  {\bibfnamefont {E.}~\bibnamefont {Granados}}, \bibinfo {author}
  {\bibfnamefont {P.}~\bibnamefont {Muggli}}, \bibinfo {author} {\bibfnamefont
  {H.}~\bibnamefont {Panuganti}}, \ and\ \bibinfo {author} {\bibfnamefont
  {G.}~\bibnamefont {{Zevi Della Porta}}},\ }\href {\doibase
  https://doi.org/10.1016/j.optlastec.2023.109921} {\bibfield  {journal}
  {\bibinfo  {journal} {Optics \& Laser Technology}\ }\textbf {\bibinfo
  {volume} {168}},\ \bibinfo {pages} {109921} (\bibinfo {year}
  {2024})}\BibitemShut {NoStop}%
\bibitem [{awa()}]{awake}%
  \BibitemOpen
  \href@noop {} {}\bibinfo {howpublished}
  {\url{https://home.cern/science/accelerators/awake}}\BibitemShut {NoStop}%
\bibitem [{\citenamefont {Caldwell}\ \emph {et~al.}(2016)\citenamefont
  {Caldwell} \emph {et~al.}}]{Caldwell2016}%
  \BibitemOpen
  \bibfield  {author} {\bibinfo {author} {\bibfnamefont {A.}~\bibnamefont
  {Caldwell}} \emph {et~al.},\ }\href {\doibase 10.1016/j.nima.2015.12.050}
  {\bibfield  {journal} {\bibinfo  {journal} {Nuclear Instruments and Methods
  \allowbreak in Physics Research Section A: Accelerators, \allowbreak
  Spectrometers, \allowbreak Detectors and Associated Equipment}\ }\textbf
  {\bibinfo {volume} {829}},\ \bibinfo {pages} {3–} (\bibinfo {year}
  {2016})},\ \bibinfo {note} {2nd European Advanced Accelerator Concepts
  Workshop - EAAC 2015}\BibitemShut {NoStop}%
\bibitem [{\citenamefont {Gschwendtner}\ \emph {et~al.}(2016)\citenamefont
  {Gschwendtner} \emph {et~al.}}]{Gschwendtner2016}%
  \BibitemOpen
  \bibfield  {author} {\bibinfo {author} {\bibfnamefont {E.}~\bibnamefont
  {Gschwendtner}} \emph {et~al.},\ }\href {\doibase 10.1016/j.nima.2016.02.026}
  {\bibfield  {journal} {\bibinfo  {journal} {Nuclear Instruments and Methods
  \allowbreak in Physics Research Section A: Accelerators, \allowbreak
  Spectrometers, \allowbreak Detectors and Associated Equipment}\ }\textbf
  {\bibinfo {volume} {829}},\ \bibinfo {pages} {76–} (\bibinfo {year}
  {2016})},\ \bibinfo {note} {2nd European Advanced Accelerator Concepts
  Workshop - EAAC 2015}\BibitemShut {NoStop}%
\bibitem [{\citenamefont {Adli}\ \emph {et~al.}(2018)\citenamefont {Adli},
  \citenamefont {Ahuja}, \citenamefont {Apsimon}, \citenamefont {Apsimon},
  \citenamefont {Bachmann}, \citenamefont {Barrientos}, \citenamefont {Batsch},
  \citenamefont {Bauche}, \citenamefont {Olsen}, \citenamefont {Bernardini}
  \emph {et~al.}}]{Adli2018}%
  \BibitemOpen
  \bibfield  {author} {\bibinfo {author} {\bibfnamefont {E.}~\bibnamefont
  {Adli}}, \bibinfo {author} {\bibfnamefont {A.}~\bibnamefont {Ahuja}},
  \bibinfo {author} {\bibfnamefont {O.}~\bibnamefont {Apsimon}}, \bibinfo
  {author} {\bibfnamefont {R.}~\bibnamefont {Apsimon}}, \bibinfo {author}
  {\bibfnamefont {A.-M.}\ \bibnamefont {Bachmann}}, \bibinfo {author}
  {\bibfnamefont {D.}~\bibnamefont {Barrientos}}, \bibinfo {author}
  {\bibfnamefont {F.}~\bibnamefont {Batsch}}, \bibinfo {author} {\bibfnamefont
  {J.}~\bibnamefont {Bauche}}, \bibinfo {author} {\bibfnamefont {V.~B.}\
  \bibnamefont {Olsen}}, \bibinfo {author} {\bibfnamefont {M.}~\bibnamefont
  {Bernardini}},  \emph {et~al.},\ }\href@noop {} {\bibfield  {journal}
  {\bibinfo  {journal} {Nature}\ }\textbf {\bibinfo {volume} {561}},\ \bibinfo
  {pages} {363} (\bibinfo {year} {2018})}\BibitemShut {NoStop}%
\bibitem [{\citenamefont {Gschwendtner}\ \emph {et~al.}(2022)\citenamefont
  {Gschwendtner}, \citenamefont {Lotov}, \citenamefont {Muggli}, \citenamefont
  {Wing} \emph {et~al.}}]{Gschwendtner2022}%
  \BibitemOpen
  \bibfield  {author} {\bibinfo {author} {\bibfnamefont {E.}~\bibnamefont
  {Gschwendtner}}, \bibinfo {author} {\bibfnamefont {K.}~\bibnamefont {Lotov}},
  \bibinfo {author} {\bibfnamefont {P.}~\bibnamefont {Muggli}}, \bibinfo
  {author} {\bibfnamefont {M.}~\bibnamefont {Wing}},  \emph {et~al.} (\bibinfo
  {collaboration} {AWAKE Collaboration}),\ }\href {\doibase
  10.3390/sym14081680} {\bibfield  {journal} {\bibinfo  {journal} {Symmetry}\
  }\textbf {\bibinfo {volume} {14}} (\bibinfo {year} {2022}),\
  10.3390/sym14081680}\BibitemShut {NoStop}%
\bibitem [{\citenamefont {Demeter}(2019)}]{Demeter2019}%
  \BibitemOpen
  \bibfield  {author} {\bibinfo {author} {\bibfnamefont {G.}~\bibnamefont
  {Demeter}},\ }\href {\doibase 10.1103/PhysRevA.99.063423} {\bibfield
  {journal} {\bibinfo  {journal} {Phys. Rev. A}\ }\textbf {\bibinfo {volume}
  {99}},\ \bibinfo {pages} {063423} (\bibinfo {year} {2019})}\BibitemShut
  {NoStop}%
\bibitem [{\citenamefont {Demeter}\ \emph {et~al.}(2021)\citenamefont
  {Demeter}, \citenamefont {Moody}, \citenamefont {Kedves}, \citenamefont
  {R\'aczkevi}, \citenamefont {Aladi}, \citenamefont {Bachmann}, \citenamefont
  {Batsch}, \citenamefont {Braunm\"uller}, \citenamefont {Djotyan},
  \citenamefont {Fedosseev}, \citenamefont {Friebel}, \citenamefont {Gessner},
  \citenamefont {Granados}, \citenamefont {Guran}, \citenamefont {H\"uther},
  \citenamefont {Lee}, \citenamefont {Martyanov}, \citenamefont {Muggli},
  \citenamefont {\"Oz}, \citenamefont {Panuganti}, \citenamefont {Verra},\ and\
  \citenamefont {Della~Porta}}]{Demeter2021}%
  \BibitemOpen
  \bibfield  {author} {\bibinfo {author} {\bibfnamefont {G.}~\bibnamefont
  {Demeter}}, \bibinfo {author} {\bibfnamefont {J.~T.}\ \bibnamefont {Moody}},
  \bibinfo {author} {\bibfnamefont {M.~A.}\ \bibnamefont {Kedves}}, \bibinfo
  {author} {\bibfnamefont {B.}~\bibnamefont {R\'aczkevi}}, \bibinfo {author}
  {\bibfnamefont {M.}~\bibnamefont {Aladi}}, \bibinfo {author} {\bibfnamefont
  {A.-M.}\ \bibnamefont {Bachmann}}, \bibinfo {author} {\bibfnamefont
  {F.}~\bibnamefont {Batsch}}, \bibinfo {author} {\bibfnamefont
  {F.}~\bibnamefont {Braunm\"uller}}, \bibinfo {author} {\bibfnamefont {G.~P.}\
  \bibnamefont {Djotyan}}, \bibinfo {author} {\bibfnamefont {V.}~\bibnamefont
  {Fedosseev}}, \bibinfo {author} {\bibfnamefont {F.}~\bibnamefont {Friebel}},
  \bibinfo {author} {\bibfnamefont {S.}~\bibnamefont {Gessner}}, \bibinfo
  {author} {\bibfnamefont {E.}~\bibnamefont {Granados}}, \bibinfo {author}
  {\bibfnamefont {E.}~\bibnamefont {Guran}}, \bibinfo {author} {\bibfnamefont
  {M.}~\bibnamefont {H\"uther}}, \bibinfo {author} {\bibfnamefont
  {V.}~\bibnamefont {Lee}}, \bibinfo {author} {\bibfnamefont {M.}~\bibnamefont
  {Martyanov}}, \bibinfo {author} {\bibfnamefont {P.}~\bibnamefont {Muggli}},
  \bibinfo {author} {\bibfnamefont {E.}~\bibnamefont {\"Oz}}, \bibinfo {author}
  {\bibfnamefont {H.}~\bibnamefont {Panuganti}}, \bibinfo {author}
  {\bibfnamefont {L.}~\bibnamefont {Verra}}, \ and\ \bibinfo {author}
  {\bibfnamefont {G.~Z.}\ \bibnamefont {Della~Porta}},\ }\href {\doibase
  10.1103/PhysRevA.104.033506} {\bibfield  {journal} {\bibinfo  {journal}
  {Phys. Rev. A}\ }\textbf {\bibinfo {volume} {104}},\ \bibinfo {pages}
  {033506} (\bibinfo {year} {2021})}\BibitemShut {NoStop}%
\bibitem [{\citenamefont {Demeter}(2025)}]{Demeter2025}%
  \BibitemOpen
  \bibfield  {author} {\bibinfo {author} {\bibfnamefont {G.}~\bibnamefont
  {Demeter}},\ }\href {\doibase
  https://doi.org/10.1016/j.optlastec.2025.114014} {\bibfield  {journal}
  {\bibinfo  {journal} {Optics \& Laser Technology}\ }\textbf {\bibinfo
  {volume} {192}},\ \bibinfo {pages} {114014} (\bibinfo {year}
  {2025})}\BibitemShut {NoStop}%
\bibitem [{\citenamefont {Bíró}\ \emph {et~al.}(2023)\citenamefont {Bíró},
  \citenamefont {Pocsai}, \citenamefont {Barna}, \citenamefont {Barnaföldi},
  \citenamefont {Moody},\ and\ \citenamefont {Demeter}}]{Biro2023}%
  \BibitemOpen
  \bibfield  {author} {\bibinfo {author} {\bibfnamefont {G.}~\bibnamefont
  {Bíró}}, \bibinfo {author} {\bibfnamefont {M.}~\bibnamefont {Pocsai}},
  \bibinfo {author} {\bibfnamefont {I.~F.}\ \bibnamefont {Barna}}, \bibinfo
  {author} {\bibfnamefont {G.~G.}\ \bibnamefont {Barnaföldi}}, \bibinfo
  {author} {\bibfnamefont {J.~T.}\ \bibnamefont {Moody}}, \ and\ \bibinfo
  {author} {\bibfnamefont {G.}~\bibnamefont {Demeter}},\ }\href {\doibase
  https://doi.org/10.1016/j.optlastec.2022.108948} {\bibfield  {journal}
  {\bibinfo  {journal} {Optics and Laser Technology}\ }\textbf {\bibinfo
  {volume} {159}},\ \bibinfo {pages} {108948} (\bibinfo {year}
  {2023})}\BibitemShut {NoStop}%
\bibitem [{\citenamefont {Schmidt}\ and\ \citenamefont
  {Lipson}(2009)}]{Schmidt2009}%
  \BibitemOpen
  \bibfield  {author} {\bibinfo {author} {\bibfnamefont {M.}~\bibnamefont
  {Schmidt}}\ and\ \bibinfo {author} {\bibfnamefont {H.}~\bibnamefont
  {Lipson}},\ }\href {\doibase 10.1126/science.1165893} {\bibfield  {journal}
  {\bibinfo  {journal} {Science}\ }\textbf {\bibinfo {volume} {324}},\ \bibinfo
  {pages} {81} (\bibinfo {year} {2009})}\BibitemShut {NoStop}%
\bibitem [{\citenamefont {Udrescu}\ and\ \citenamefont
  {Tegmark}(2020)}]{Udrescu2020}%
  \BibitemOpen
  \bibfield  {author} {\bibinfo {author} {\bibfnamefont {S.-M.}\ \bibnamefont
  {Udrescu}}\ and\ \bibinfo {author} {\bibfnamefont {M.}~\bibnamefont
  {Tegmark}},\ }\href {\doibase 10.1126/sciadv.aay2631} {\bibfield  {journal}
  {\bibinfo  {journal} {Science Advances}\ }\textbf {\bibinfo {volume} {6}},\
  \bibinfo {pages} {eaay2631} (\bibinfo {year} {2020})}\BibitemShut {NoStop}%
\bibitem [{\citenamefont {Cranmer}(2023)}]{Cranmer2023}%
  \BibitemOpen
  \bibfield  {author} {\bibinfo {author} {\bibfnamefont {M.}~\bibnamefont
  {Cranmer}},\ }\href {\doibase 10.48550/arXiv.2305.01582} {\bibfield
  {journal} {\bibinfo  {journal} {arXiv:2305.01582}\ } (\bibinfo {year}
  {2023}),\ 10.48550/arXiv.2305.01582}\BibitemShut {NoStop}%
\bibitem [{\citenamefont {van Lange}\ \emph {et~al.}(2020)\citenamefont {van
  Lange}, \citenamefont {van~der Straten},\ and\ \citenamefont {van
  Oosten}}]{vanLange2020}%
  \BibitemOpen
  \bibfield  {author} {\bibinfo {author} {\bibfnamefont {A.~J.}\ \bibnamefont
  {van Lange}}, \bibinfo {author} {\bibfnamefont {P.}~\bibnamefont {van~der
  Straten}}, \ and\ \bibinfo {author} {\bibfnamefont {D.}~\bibnamefont {van
  Oosten}},\ }\href {\doibase 10.1088/1361-6455/ab7fc2} {\bibfield  {journal}
  {\bibinfo  {journal} {Journal of Physics B: Atomic, Molecular and Optical
  Physics}\ }\textbf {\bibinfo {volume} {53}},\ \bibinfo {pages} {125402}
  (\bibinfo {year} {2020})}\BibitemShut {NoStop}%
\bibitem [{\citenamefont {Saleh}\ and\ \citenamefont
  {Teich}(2019)}]{SalehTeich}%
  \BibitemOpen
  \bibfield  {author} {\bibinfo {author} {\bibfnamefont {B.~E.~A.}\
  \bibnamefont {Saleh}}\ and\ \bibinfo {author} {\bibfnamefont {M.~C.}\
  \bibnamefont {Teich}},\ }\href@noop {} {\emph {\bibinfo {title} {Fundamentals
  of Photonics}}}\ (\bibinfo  {publisher} {John Wiley and Sons, Inc.},\
  \bibinfo {year} {2019})\BibitemShut {NoStop}%
\bibitem [{\citenamefont {Adli}\ \emph {et~al.}(2019)\citenamefont {Adli} \emph
  {et~al.}}]{Adli2019v2}%
  \BibitemOpen
  \bibfield  {author} {\bibinfo {author} {\bibfnamefont {E.}~\bibnamefont
  {Adli}} \emph {et~al.} (\bibinfo {collaboration} {AWAKE Collaboration}),\
  }\href {\doibase 10.1103/PhysRevLett.122.054802} {\bibfield  {journal}
  {\bibinfo  {journal} {Phys. Rev. Lett.}\ }\textbf {\bibinfo {volume} {122}},\
  \bibinfo {pages} {054802} (\bibinfo {year} {2019})}\BibitemShut {NoStop}%
\bibitem [{\citenamefont {Demeter}()}]{Demeter_tbp}%
  \BibitemOpen
  \bibfield  {author} {\bibinfo {author} {\bibfnamefont {G.}~\bibnamefont
  {Demeter}},\ }\href@noop {} {}\bibinfo {note} {To be published}\BibitemShut
  {NoStop}%
\bibitem [{\citenamefont {H{\'{e}}der}\ \emph {et~al.}(2022)\citenamefont
  {H{\'{e}}der}, \citenamefont {Rig{\'{o}}}, \citenamefont {Medgyesi},
  \citenamefont {Lovas}, \citenamefont {Tenczer}, \citenamefont
  {T{\"o}r{\"o}k}, \citenamefont {Farkas}, \citenamefont {Em{\H{o}}di},
  \citenamefont {Kadlecsik}, \citenamefont {Mez{\H{o}}}, \citenamefont
  {Pint{\'{e}}r},\ and\ \citenamefont {Kacsuk}}]{Heder2022}%
  \BibitemOpen
  \bibfield  {author} {\bibinfo {author} {\bibfnamefont {M.}~\bibnamefont
  {H{\'{e}}der}}, \bibinfo {author} {\bibfnamefont {E.}~\bibnamefont
  {Rig{\'{o}}}}, \bibinfo {author} {\bibfnamefont {D.}~\bibnamefont
  {Medgyesi}}, \bibinfo {author} {\bibfnamefont {R.}~\bibnamefont {Lovas}},
  \bibinfo {author} {\bibfnamefont {S.}~\bibnamefont {Tenczer}}, \bibinfo
  {author} {\bibfnamefont {F.}~\bibnamefont {T{\"o}r{\"o}k}}, \bibinfo {author}
  {\bibfnamefont {A.}~\bibnamefont {Farkas}}, \bibinfo {author} {\bibfnamefont
  {M.}~\bibnamefont {Em{\H{o}}di}}, \bibinfo {author} {\bibfnamefont
  {J.}~\bibnamefont {Kadlecsik}}, \bibinfo {author} {\bibfnamefont
  {G.}~\bibnamefont {Mez{\H{o}}}}, \bibinfo {author} {\bibfnamefont
  {{\'{A}}.}~\bibnamefont {Pint{\'{e}}r}}, \ and\ \bibinfo {author}
  {\bibfnamefont {P.}~\bibnamefont {Kacsuk}},\ }\href {\doibase
  10.22503/inftars.xxii.2022.2.8} {\bibfield  {journal} {\bibinfo  {journal}
  {Inform{\'{a}}ci{\'{o}}s T{\'{a}}rsadalom}\ }\textbf {\bibinfo {volume}
  {22}},\ \bibinfo {pages} {128} (\bibinfo {year} {2022})}\BibitemShut
  {NoStop}%
\end{thebibliography}%

\bibliographystyle{apsrev4-1}


\end{document}